\documentclass[prl,showpacs,twocolumn]{revtex4-1} 
\usepackage{graphicx}
\usepackage{amsmath}
\usepackage{amssymb}
\usepackage{bm}

\usepackage{mathpazo}

\newcommand{\be}{\begin{eqnarray}}
\newcommand{\ee}{\end{eqnarray}}
\newcommand{\nn}{\nonumber}
\newcommand{\bn}{\begin{enumerate}}
\newcommand{\en}{\end{enumerate}}

\def\IC{\mathbb{C}}

\def\IR{\mathbb{R}}

\def\CA{{\cal A}}
\def\CB{{\cal B}}
\def\CC{{\cal C}}

\def\CJ{{\cal J}}

\def\CL{{\cal L}}

\def\CN{{\cal N}}

\def\CR{{\cal R}}

\def\CZ{{\cal Z}}

\def\a{\alpha}
\def\b{\beta}

\def\d{\delta}
\def\e{\epsilon}

\def\z{\zeta}
\def\th{\theta}

\def\l{\lambda}
\def\m{\mu}

\def\L{\Lambda}

\def\thalf{{\textstyle \frac{1}{2}}}

\def\det{{\rm det}}

\def\da{{\dot{\a}}}

\def\jmath{{j}}

\begin{document}

\title{Yangian Invariant Scattering Amplitudes in Supersymmetric Chern-Simons Theory}
\author{Sangmin \surname{Lee}}
\affiliation{Department of Physics, 
University of Seoul, Seoul 130-743, Korea}

\begin{abstract}
We propose a generating function for scattering amplitudes 
of $\CN=6$ super-Chern-Simons theory 
which parallels a recent work on $\CN=4$ super-Yang-Mills theory by Arkani-Hamed et al. 
Our result suggests 
that the scattering amplitudes of 
the super-Chern-Simons theory 
exhibit Yangian invariance. 

\end{abstract}


\maketitle


\paragraph{Introduction.}

The last decade has seen remarkable advances 
in novel methods for computing scattering amplitudes in 
perturbative Yang-Mills theories; see, e.g., \cite{Bern:2007dw,Wolf:2010av} for recent reviews.  
While the new techniques are applicable for many theories including QCD, 
the $\CN=4$ super-Yang-Mills theory (SYM$_4$) 
has proved to be the richest 
testing ground for new theoretical ideas.    

Recently, Arkani-Hamed {\it et al.} \cite{ArkaniHamed:2009dn} 
proposed a remarkably simple reformulation of the scattering amplitudes 
of planar SYM$_4$. They presented a ``generating function for scattering amplitudes'' named $\CL_{n,k}$ in the form of a matrix-valued contour integral. 
With a suitable choice of the integration contour, 
$\CL_{n,k}$ was conjectured to capture the leading singularities 
associated to $n$-point, $k$ negative helicity amplitudes.
In particular, it has been proven that $\CL_{n,k}$ reproduces 
all tree-level amplitudes $\CA^{\rm tree}_{n,k}$ \cite{Bourjaily:2010kw}. 
The applicability of the formulation of \cite{ArkaniHamed:2009dn} 
to loop amplitudes is less clear; see 
\cite{ArkaniHamed:2010kv} for the most recent progress. 

One of the most striking features of SYM$_4$, 
which looks mysterious from traditional points of view 
but becomes transparent in the new formulation, 
is the so-called dual superconformal symmetry \cite{Alday:2007hr,Drummond:2008vq,Berkovits:2008ic,Beisert:2008iq}.
The original and dual superconformal symmetry of SYM$_4$ 
together generate a Yangian symmetry \cite{Drummond:2009fd}, 
which introduces elements of integrability 
into the scattering amplitudes of SYM$_4$ 
in the planar sector \cite{Beisert:2010jq} . 

It is clearly interesting to see if the new formulation of \cite{ArkaniHamed:2009dn} 
with a built-in Yangian symmetry can be applied to other field theories. 
One strong candidate is 
the three dimensional $\CN=6$ super-Chern-Simons theory (SCS$_6$) 
constructed in \cite{Aharony:2008ug,scs4,scs6}. 
This theory is special because of two common features 
it shares with SYM$_4$; 
it has a string theory dual in the sense of \cite{adscft} 
and its superconformal algebra admits a simple extension 
to Yangian algebra as explained in \cite{Drummond:2009fd}.

Preliminary studies on scattering amplitudes of SCS$_6$
\cite{Agarwal:2008pu,Bargheer:2010hn,Huang:2010rn} 
reported results in favor of Yangian symmetry. 
These findings came as a surprise since 
earlier attempts had come short of realizing dual superconformal symmetry 
in the string theory dual \cite{Adam:2009kt, Grassi:2009yj}. 
Dual superconformal symmetry would imply a Yangian symmetry,   
although the converse does not hold. 
Further studies are required to see 
whether the Yangian symmetry of SCS$_6$ 
extends to all tree-level amplitudes 
and, if so, whether it originates from some 
dual superconformal symmetry.

The aim of this Letter is to provide further support 
for the relevance of Yangian symmetry in SCS$_6$. 
Generalizing the approach of \cite{ArkaniHamed:2009dn},  
we present a generating function for scattering amplitudes of SCS$_6$ -- $\CL_{2k}$ in eq. (\ref{3d-grass}) -- in the same sense as 
$\CL_{n,k}$ of SYM$_4$ and give a formal proof of its Yangian invariance 
for all $k$. 

We begin with a quick review of Witten's twistor formulation \cite{Witten:2003nn} on which $\CL_{n,k}$ of \cite{ArkaniHamed:2009dn} is based,
and explain how it should be generalized to three dimensions.  
Using a supersymmetric version of the three dimensional twistor, 
we write down the generating function $\CL_{2k}$ for SCS$_6$ 
and study its properties. We verify superconformal invariance, 
cyclic symmetry and Yangian invariance,  
and also show that it reproduces some known tree-level amplitudes. 
We conclude with a discussion on dual superconformal symmetry 
and directions for future works.  

\paragraph{Twistor in 4D vs 3D.} 

In four dimensions, a null momentum can be written as a bi-spinor $p^{\a\da} = \l^\a \bar{\l}^{\da}$. 
A standard way to introduce twistors \cite{Witten:2003nn} 
is to take a Fourier transform of the plane wave $e^{ip\cdot x}$ 
with respect to one of the two spinors:
\be
\int e^{i x_{\a\da} \l^\a \bar{\l}^\da} e^{-i\bar{\m}_\a \l^\a} d^2\l 
\;\propto\; \d(\bar{\m}_\a - x_{\a\da} \bar{\l}^\da) \,.
\ee
The delta function enforces the defining equation for 
the twistor variables $(\bar{\m}_\a, \bar{\l}^\da)$.
Equivalently, we can regard 
$\bar{\mu}_\a = -i (\partial/\partial \l^\a)$ as a ``momentum operator" 
acting on ``wave-functions" in the $(\l,\bar{\l})$-space 
and re-interpret the twistor equation as a wave equation,  
\be
(\bar{\m}_\a - x_{\a\da} \bar{\l}^\da) \exp(i x_{\a\da} \l^\a \bar{\l}^\da) = 0 \,.
\ee

In three dimensions, the bi-spinor notation involves a single spinor, $p^{\a\b} = \l^\a\l^\b$. 
Introducing the operator $\m_\a = i(\partial/ \partial \l^\a)$, 
we can again realize the three dimensional twistor equation \cite{Hitchin:1982gh,Chiou:2005jn}
as a wave equation:
\be
(\m_\a - x_{\a\b}\l^\b) \exp\left(-\frac{i}{2} x_{\a\b} \l^\a \l^\b\right) 
= 0 \,.
\ee 

Drawing an analogy from quantum mechanics, 
we note that the $n$-point amplitude can be treated as a wave function 
in $2n$ ``coordinate" variables $\{\l_i^\a\}$ $(i=1,\cdots,n)$, whereas the 
conformal symmetry 
${\rm SO}(2,3) \simeq {\rm Sp}(4,\IR)$ acts linearly on the $4n$ dimensional ``phase space" parametrized by $\{Z_i^A= (\l_i^\a,\mu_{i\a})\}$.

\paragraph{Super-twistor.} 
The on-shell superfield for SCS$_6$ involves 
three fermionic coordinates $\eta^I$ in addition to $\l^\a$ \cite{Bargheer:2010hn}. The particle and antiparticle superfields 
take the form
\be
&&\Phi = \phi^4 + \eta^I \psi_I + \thalf \e_{IJK} \eta^I\eta^J \phi^K 
+ \tfrac{1}{6} \e_{IJK} \eta^I \eta^J \eta^K \psi_4 \,,
\nn \\
&&\bar{\Phi} = \bar{\psi}^4 + \eta^I \bar{\phi}_I + 
\thalf \e_{IJK} \eta^I\eta^J \bar{\psi}^K 
+ \tfrac{1}{6} \e_{IJK} \eta^I \eta^J \eta^K \bar{\phi}_4 \,,
\label{sfield}
\ee
where the scalars $\phi$ and fermions $\psi$ are all understood as 
functions of the momentum spinor $\l$. 
The SO(6) $R$-symmetry acting on the $\CN=6$ supercharges 
are realized by $\eta^I$ and their conjugates $\z_I = \partial/\partial \eta^I$ 
through the oscillator algebra, 
\be
\{ \eta^I , \z_J \} = \d^I_J 
\;\;\;\;\; 
(I,J=1,2,3)  \,.
\ee

Note that the quantum mechanics analogy introduced 
above remains valid even after including fermions;
while the super-amplitude $\CA(\L)$ 
can be regarded as a wave-function in the ``half super-twistor" variables $\L_i = (\l^\a, \eta^I)_i$, 
the generators of the full superconformal symmetry OSp$(6|4)$ are represented 
by quadratic products of the ``full super-twistor" variables 
\be
\CZ^\CA_i = (\l^\a, \m_{\a}; \eta^I, \zeta_{I})_i \sim (\L , \partial/\partial{\L})_i 
\ee
to be interpreted as operators acting on $\CA(\L)$.    
In this formulation, only the U$(1,1|3) \subset {\rm OSp}(6|4)$ 
acts linearly on $\L$ with generators of the form 
$(\L\partial/\partial \L)$.  
The rest of the generators act, schematically, either as multiplications $(\L \L)$ or 
as second order derivatives $(\partial^2/\partial \L \partial \L)$.
Although both $\l$ and $\eta$ are subject to reality conditions, 
we will loosely treat them as complex variables $\L \in \IC^{2|3}$ 
in what follows. 

\paragraph{Amplitudes of SCS$_6$.}

The superfields $\Phi$ and $\bar{\Phi}$ in  
(\ref{sfield}) transform in mutually complex conjugate representations 
of the gauge group; a prime example is ${\rm U}(N)\times {\rm U}(N)$ gauge group 
with $\Phi$ transforming in $(\mathbf{N},\overline{\mathbf{N}})$ and $\bar{\Phi}$ in 
$(\overline{\mathbf{N}},\mathbf{N})$. 
Barring the possibility of a ``baryonic" vertex such as $\det(\Phi)$ 
which scale as $\Phi^N$, the non-vanishing amplitudes 
must carry equal number of $\Phi$ and $\bar{\Phi}$.
Moreover, one can define color-ordered amplitudes such that 
the external legs alternate between $\Phi$ and $\bar{\Phi}$ \cite{Bargheer:2010hn}. 
In summary, we are interested in the $n(=2k)$-point color-ordered super-amplitudes
\be
\CA_{n=2k}(\L) = \CA_{2k}(\L_1, \L_2, \cdots, \L_{2k}), 
\ee
where by convention we associate $\L_{\rm odd/even}$ to $\bar{\Phi}$/$\Phi$ 
(opposite to the convention of \cite{Bargheer:2010hn}).
Because $\Phi$ and $\bar{\Phi}$ carry opposite statistics, 
$\CA_{2k}$ acquires a factor of $(-1)^{k-1}$ upon cyclic permutation 
by two sites \cite{Bargheer:2010hn}.
The component amplitudes can be read off from the 
super-amplitude as the coefficients of various monomials of $\eta^I_i$. 
They are rational functions of Lorentz invariant products of the 
momentum spinors, 
$\langle i j \rangle \equiv \e_{\a\b} \l_i^\a \l_j^\b$. 

\paragraph{The generating function.}

Our proposal for the generating function for the 
$n(=2k)$-point amplitude is 
\be
\CL_{2k}(\L) = 
\int  \frac{d^{k\times 2k} C}{{\rm vol}[{\rm GL}(k)]} \frac{\d^{\frac{k(k+1)}{2}}(C\cdot C^T)\, \d^{2k|3k}(C\cdot \L)}
{M_1 M_2 \cdots M_k}
\,.
\label{3d-grass}
\ee
This form of $\CL_{2k}$ was partly motivated by the formal similarity 
noted in \cite{Bargheer:2010hn}
between $\CA_{2k}$ of SCS$_6$ and $\CA_{2k,k}$ of SYM$_4$. 
As will become gradually clearer, 
both the similarity and the difference between $\CL_{n,k}$ 
of \cite{ArkaniHamed:2009dn} and $\CL_{2k}$ here can be traced back to 
the structure of the momentum-spinor: $p^{\a\da} = \l^\a \bar{\l}^{\da}$ 
in four dimensions and $p^{\a\b} = \l^\a \l^\b$ in three dimensions.

The integration variable $C$ is a $(k\times 2k)$ matrix. 
The dot products denote $(C\cdot C^T)_{mn} = C_{mi}C_{ni}$, $(C\cdot \L)_m = C_{mi}\L_i$.
$M_i$ represents the $i$-th minor of $C$ defined by
\be
M_i = \e^{m_1 \cdots m_k} C_{m_1 (i)} C_{m_2 (i+1)} \cdots C_{m_k (i+k-1)} \,.
\ee
The measure $d^{k\times 2k}C$ is covariant under a ${\rm GL}(k)\times {\rm GL}(2k)$ group  
action on the left/right. 
The vol$[{\rm GL}(k)]^{-1}$ factor is a reminder that the ${\rm GL}(k)$-left action is an exact symmetry of the integral and should be ``gauge fixed".
The ${\rm GL}(2k)$-right would-be symmetry is reduced to ${\rm O}(2k)$ by $\d(C\cdot C^T)$, which is in turn broken spontaneously by $\d(C\cdot\L)$ and 
explicitly by the denominator.

The cyclic symmetry of (\ref{3d-grass}) is  
obscured by the presence of only $k$ out of $2k$ minors. 
However, one can use the constraint $C\cdot C^T=0$ to show that  
\be
M_i M_{i+1} = (-1)^{k-1} M_{i+k} M_{i+1+k} \,.
\ee
Thus $\CL_{2k}$ transforms in the same way as $\CA_{2k}$
 under cyclic permutation by two sites.

The net number of integration variables in (\ref{3d-grass}) 
can be counted as follows (cf. \cite{ArkaniHamed:2009dn}). 
Starting from $2k^2$ elements of $C$, subtracting $k^2$ for the ${\rm GL}(k)$ gauge fixing and 
$2k$ for the bosonic delta functions, and pulling out the overall momentum 
conserving delta function, we are left with 
\be
\label{counting}
2k^2 - k^2 - \frac{k(k+1)}{2} - 2k +3 = \frac{(k-2)(k-3)}{2} \,. 
\ee

\paragraph{Superconformal invariance.} 
To begin with, note that $\CL_{2k}$ has degree $-2k$ in $\lambda$ and $+3k$ in $\eta$ 
in agreement with the degree counting for $\CA_{2k}$ from Feynman diagrams \cite{Bargheer:2010hn}. 
To verify superconformal invariance in the half super-twistor notation, 
we need to consider three cases separately. 
The $\d(C\cdot \L)$ factor is manifestly invariant under the linearly realized 
${\rm U}(1,1|3)$ subgroup. 
The two-derivative generators $(\partial^2/\partial \L \partial \L)$  acting on $\d^{2k|3k}(C\cdot \L)$ 
produces $C\cdot C^T$ to be annihilated by $\d(C\cdot C^T)$.  
To see the invariance under the multiplication generators $(\L\L)$, 
note that the constraint $C\cdot C^T = 0$ generically defines $k$ linearly independent 
null vectors in $\IC^{2k}$. One can construct another $(k\times 2k)$ matrix \
$\widehat{C}$ composed of $k$ dual null-vectors in $\IC^{2k}$ satisfying
\be
\widehat{C} \cdot \widehat{C}^T =0 \,,
\;\;\; 
C\cdot \widehat{C}^T =  I_{k\times k} \,,
\ee
and use the completeness relation for $C$ and $\widehat{C}$ to write
\be
\L^T \cdot \L = \L^T \cdot (C^T \widehat{C} + \widehat{C}^T C) \cdot \L \,,
\ee
which is annihilated by the $\d(C\cdot \L)$ factor.

\paragraph{Reproducing known amplitudes.}

Only 4- and 6-point amplitudes of SCS$_6$ are available in the literature so far  
\cite{Agarwal:2008pu,Bargheer:2010hn,Huang:2010rn}. 
In both cases, it is straightforward to show that $\CL_{2k}$ 
reproduces the amplitudes mainly because there is no integral to do according to the counting 
(\ref{counting}). 

Here, we only discuss the 4-point amplitude and refer the reader 
to \cite{wip} for the 6-point result. 
For simplicity, we begin with the 4-scalar component amplitude \cite{Bargheer:2010hn},  
\be
A_{4\phi} =  
\frac{\langle 13 \rangle^3}{\langle 14 \rangle\langle 43 \rangle} 
\d^{(3)}(\l_r\l_r + \l_{\bar{s}} \l_{\bar{s}}) \,,
\ee
where we divided the particle indices $\{i=1,\cdots,4\}$ 
into $\{r=1,3\}$ and $\{\bar{s}=2,4\}$. Consider
the partial Fourier transform, 
\be
\widehat{A}_{4\phi}(\l_r,\m_{\bar{s}}) = \int A_{4\phi}(\l_r,\l_{\bar{s}}) 
e^{-i\m_{\bar{s}} \l_{\bar{s}}} d^4\l_{\bar{s}} \,, 
\ee
and introduce the ``link matrices" (cf. \cite{ArkaniHamed:2009si}) 
defined by
\be
\l_{\bar{s}} = - c_{r\bar{s}} \l_r \,.
\ee
After the change of variable from $\l_{\bar{s}}^\a$ 
to $c_{r\bar{s}}$, we obtain (up to an overall coefficient)
\be
\label{4pt-link}
\widehat{A}_{4\phi}(\l_r,\m_{\bar{s}}) = \int \frac{d^4 c_{r\bar{s}}
\d^{(3)}(\d_{rp} + c_{r \bar{s}} c_{p \bar{s}})}{c_{14}c_{34}} e^{i c_{r\bar{s}} \l_r \m_{\bar{s}}} \,.
\ee
Taking the inverse Fourier transform back to $A_{4\phi}(\l_r,\l_{\bar{s}})$ 
and reinstating the fermions, we recognize the final result as a gauge fixed version of $\CL_{2k}$ with 
\be
C=
\begin{pmatrix}
c_{12} & 1 & c_{32} & 0 \\
c_{14} & 0 & c_{34} & 1  
\end{pmatrix} \,.
\ee

\paragraph{Integrability via Yangian symmetry.} 

The original superconformal invariance alone 
is far from sufficient 
to determine the amplitude uniquely. 
For instance, we can multiply $\CL_{2k}$  
by an arbitrary function $f(\CZ_i \wedge \CZ_j)$ without 
breaking superconformal invariance, where the product $\CZ_i \wedge \CZ_j$ 
is defined by the ${\rm OSp}(6|4)$-invariant metric. 

In four dimensions, under mild assumptions, $\CL_{n,k}$ was proven to 
be the unique Yangian invariants \cite{Drummond:2010qh,Drummond:2010uq,Korchemsky:2010ut}.  
Encouraged by the Yangian invariance of 4- and 6-point 
amplitudes \cite{Bargheer:2010hn} 
and the fact that $\CL_{2k}$ reproduces them, 
we now move on to examine the Yangian invariance of $\CL_{2k}$ for all $k$.  
We will show that $\CL_{2k}$ is annihilated by the level one Yangian generators, which together with the superconformal invariance 
guarantees the full Yangian invariance \cite{Drummond:2009fd}. 
The uniqueness problem is left for a future work. 

We mostly follow the methods developed in \cite{Drummond:2010qh} to prove the Yangian invariance of $\CL_{n,k}$.  
As shown in \cite{Drummond:2009fd,Bargheer:2010hn}, the level one Yangian generators can be written in the bilinear form, 
\be
\CJ^{\CA}{}_{\CB} = \sum_{i<j} (-1)^{\CC} \left(J_i^{\CA}{}_{\CC} 
J_j^{\CC}{}_{\CB} -J_j^{\CA}{}_{\CC} 
J_i^{\CC}{}_{\CB} \right) \,,
\ee 
where $J_{i}{}^\CA{}_\CB$ are the superconformal generators acting on the $i$-th particle. In terms of the full super-twistors $\CZ^\CA_i$, 
the generators  
can be written as
\be
\CJ^{\CA}{}_{\CB} =  \sum_{i<j} \left[ 
\CZ^\CA_i \CZ_{\CB j} \CZ^\CC_i \CZ_{\CC j} -i \CZ^\CA_{i}\CZ_{\CB i}  
- (i \leftrightarrow j) \right] \,.
\label{yang2}
\ee 
The key insight we adopt from \cite{Drummond:2010qh} is that 
$\CZ^{\CC}_i \CZ_{\CC j}$ generates 
an ${\rm O}(2k)$ action on $\{ \L_i\}$. 
Using the covariance of $\d^{2k|3k}(C\cdot\L)$, 
we can trade it with an inverse ${\rm O}(2k)$ action on the matrix $C$. 
In other words, we can replace $\CZ^{\CC}_i \CZ_{\CC j}$ 
by $O_{ij} = i \sum_{m=1}^{k} (C_{m i} \partial/\partial C_{m j} - C_{m j} \partial/\partial C_{m i})$. 
The factors $dC$ and $\d(C\cdot C^T)$ are invariant under the ${\rm O}(2k)$ action, 
so we can do an integration by parts to make $O_{ij}$ act on the denominator. 
Then, following essentially the same steps as in \cite{Drummond:2010qh}, 
we can show that the quartic and quadratic terms in (\ref{yang2}) 
acting on $\CL_{2k}$ cancel each other.

\paragraph{Dual superconformal symmetry and momentum-twistor.} 
In four dimensions, there is an alternative way to prove the Yangian invariance 
directly through its relation to dual superconformal symmetry \cite{Drummond:2009fd}. 
It was shown in \cite{ArkaniHamed:2009vw} 
that after a suitable change of variables, $\CL_{n,k}$ can be 
rewritten as $\CL_{n,k} = \CA^{\rm MHV}_n \times \CR_{n,k}$, 
where $\CA^{\rm MHV}_n$ is the $n$-point maximally helicity violating (MHV) 
amplitude and $\CR_{n,k}$ is another integral formula 
with manifest dual superconformal invariance \cite{Mason:2009qx}. 
The change of variable to go from $\CL_{n,k}$ to $\CR_{n,k}$ 
can be interpreted in terms of the ``momentum twistor" introduced in \cite{Hodges:2009hk}.

In our case, it is straightforward to make a change of variable 
similar to that of \cite{ArkaniHamed:2009vw} to obtain
\be
\label{dual-grass}
\CL_{2k} = \frac{\d^{3}(P)\d^6(Q)}
{[\langle 12 \rangle \cdots \langle 2k-1\, 2k \rangle \langle 2k\, 1 \rangle]^{1/2}} 
\times \CR_{2k} \,.
\ee
This factorization was noted previously in \cite{Bargheer:2010hn}.
For the bosonic variables, the notion of momentum twistor continues to hold 
with little modification. But, we have not 
found a satisfactory interpretation 
for the new fermionic coordinates that enter $\CR_{2k}$ 
in terms of a ``momentum super-twistor''.

The relation between dual superconformal symmetry and 
Yangian symmetry in SCS$_6$ has been elucidated 
in a very recent paper \cite{Arthur}. The dual superconformal generators 
are 
naturally defined in the ``dual space" (cf. \cite{Drummond:2008vq}), 
\be
x_i^{\a\b} - x_{i+1}^{\a\b} = \l_i^\a \l_i^\b \,,
\;\;\;
\th_i^{I\a} - \th_{i+1}^{I\a} = \l_i^\a \eta_i^I \,. 
\ee 
According to \cite{Arthur}, dual superconformal symmetry 
of SCS$_6$ requires some additional bosonic coordinates in the dual space, 
which account for the missing piece in previous search \cite{Adam:2009kt, Grassi:2009yj} for dual superconformal symmetry via ``fermionic T-duality"  in string theory \cite{Berkovits:2008ic,Beisert:2008iq}.
How the results of \cite{Arthur} 
may relate to the momentum super-twistor is an interesting open problem.

\paragraph{Outlook.} 
Although the results of this Letter are quite suggestive, 
much work remains to be done 
to establish the connection between the generating function 
$\CL_{2k}$ and the amplitudes $\CA_{2k}$ 
to the same extent as their four dimensional counterparts.  
Among other things, a precise prescription for 
the integration contour will be needed to 
deduce recursion relations 
analogous to \cite{Cachazo:2004kj,Britto:2005fq} from $\CL_{2k}$, 
which in turn could be related to the usual  
perturbation theory in terms of Feynman diagrams.  
A completely rigorous proof of Yangian invariance 
\cite{Bargheer:2009qu} taking account of anomalies in collinear limits 
\cite{Cachazo:2004by} will 
also rely on the correct choice of contours. 
A variant of the geometric picture based on Grassmannian explained in \cite{ArkaniHamed:2009dn} will be very useful in solving the contour problem. 
Some of these issues are currently under investigation and will be reported in \cite{wip}.

\begin{acknowledgments}
We thank the organizers of 
the Carg\`ese Summer School on String Theory: Formal Developments and Applications (June 21 - July 3, 2010), 
where the work was initiated with inspiration from  
the lectures of N. Arkani-Hamed and Z. Bern. 
We are especially indebted to A. Lipstein and Y.-t. Huang 
for many illuminating discussions and for sharing their work \cite{Arthur} \
prior to publication. 
We also benefitted from discussions with D. Bak and H. Johansson.
This work was supported in part by National 
Research Foundation of Korea (NRF) Grants 
No. 2007-331-C00073, 2009-0072755 and 2009-0084601. 
\end{acknowledgments}

\end{document}